\documentstyle[12pt,openbib]{article}
\begin{document}
\title{\bf How reliable is the mean-field Nuclear Matter description for 
supporting Chiral Effective Lagrangians?}
\vskip .3cm
\author{A. Delfino $^{1}$ \thanks{Partially supported by CNPq of 
Brasil},
 M. Malheiro $^{1}$ \thanks{Partially
supported by CNPq and CAPES of Brasil, Present address: Department of 
Physics, University of Maryland, College Park, Maryland 20742-4111, USA},  
and T. Frederico $^{2}$ 
\thanks{Partially supported by CNPq of Brasil}\\ \\ $^{1}$ Instituto de 
F\'\i sica, Universidade Federal Fluminense, \\ 24210-340, Niter\'oi,
R. J., Brasil
\\${^2}$ Instituto Tecnol\'ogico da Aeron\'autica\\
12331 S\~ao Jos\'e dos Campos, S\~ao Paulo, Brasil}
\vspace{.3 cm} 
\date{\today}
\maketitle
{\it Abstract} :
        The link between non-linear chiral effective Lagrangians 
and the Walecka model description of bulk nuclear matter [1]
is questioned. This fact is by itself
due to the Mean Field Approximation (MFA) which in nuclear mater 
makes the picture of a nucleon-nucleon interaction 
based on scalar(vector) meson exchange,  
equivalent to the description of a nuclear matter based on
attractive and repulsive contact interactions. We present
a linear chiral model where this link between the Walecka model and
an underlying to chiral symmetry realization still holds, due to MFA.   
\newpage
Very recently a chiral effective Lagrangian description of nuclear
matter has been suggested [1]. Such a proposal was based 
on adding a four-fermion Lagrangian term to 
a general Lagrangian, describing the interactions of pions and
nucleons in which the spontaneously broken SU(2) x SU(2) chiral 
symmetry is non-linearly realized. In short, the constructed non-
linear chiral Lagrangian is
\\
$$ {\cal L}_{NLC} = {\cal L}_{Weinberg}\,\,+\,\,{\cal L}_{4-N},\,\,\,\,
\eqno(1)$$
\\
where
\\
$$ {\cal L}_{4-N} =\,\,\, \frac{1}{2}\,G_{s}^{2} (\bar\psi \psi)^2\,\,\, - 
\,\, \frac{1}{2} \,G_{v}^2  
(\bar \psi \gamma_{\mu }\psi)^2,\,\,\,,\eqno(2)$$
\\
It turned out from that study 
that, at the mean-field level, the equations 
of state for this model are nothing but those obtained from the 
standard Walecka model [2] for infinite nuclear matter. 
Therefore, one gets the picture of 
a connection between a non-linear chiral Lagrangian and a baryonic
phenomenological model. At the level of mean-field this is correct. 
In order to understand how far we can proceed beyond this picture
it is important to see where all this comes from.  

To study the properties of hadronic matter, Walecka [2] proposed
a simple renormalizable model based on field theory, which is often
referred to as Quantum Hadrodynamics (QHD). In this model nucleons
interact through the exchange of $\sigma$ and $\omega$ mesons, with 
$\sigma$ simulating medium range attraction and $\omega$
simulating short-range repulsion. The usual
approach to solve this model is the mean field approximation (MFA), in
which the meson fields are replaced by their expectation values. 
A curious aspect of this model is that the scalar (vector) 
$m_\sigma$ ($m_\omega$) masses and $g_\sigma$ ($g_\omega$) coupling
constants in the equations of state for infinite 
nuclear matter are eliminated in favour of\, 
$C_{\sigma}^2 = g_{\sigma}^2M^2/m_{\sigma}^2$\,  and
\, $C_{\omega}^2 = g_{\omega}^2M^2/m_{\omega}^2$ . Since  
$C_{\sigma}^2$ and  $C_{\omega}^2$ are fitted to reproduce
the nuclear matter bulk properties, values of $m_{\sigma}$ and
$m_{\omega}$ become irrelevant. It means that this model cannot
distinguish between arbitrary values for the mesonic masses. 
Such arbitrariness can be extended to 
include infinite mesonic masses. If this is the case ,
one is led to the situation where the interaction between
the nucleons is zero-range. By using attractive and 
repulsive contact interactions it is easy to show that the above 
conjecture is correct and therefore, a Zero Range Model (ZRM) is
equivalent to the Walecka model for infinite nuclear matter
at the MFA level. It means that, if we start with 
\\
$$ {\cal L}_{ZRM} = \bar \psi i \gamma_{\mu} \partial^{\mu } \psi -
\bar \psi M \psi  +\,\frac{1}{2}G_{s}^{2} (\bar\psi \psi)^2 - \,
\frac{1}{2}G_{v}^2  
(\bar \psi \gamma_{\mu }\psi)^2,\,\,\,,\eqno(3)$$
\\
where ${\psi}$'s are baryonic fields, we arrive 
at the same equations of state of the Walecka model if one just
multiplies by $M^2$ the coupling constants.   
This strange picture is therefore constructed from MFA itself 
applied for
infinite nuclear matter, and has nothing strictly to do with chiral 
invariance symmetry. The Lagrangian given by Eq.(1) and the
 Lagrangian given by Eq.(3) lead to the same equations of state
obtained from the Walecka Lagrangian at MFA level for infinite
nuclear matter. They do not differ at the MFA level 
when surfaces effects are not included.     
Further distorted pictures could be obtained also
for the self-coupling nonlinear  
\,$\sigma\,-\,\omega$ model if one adds terms such as 
\,\,\,$a(\bar\psi \psi)^3$\,\,\,\, and\,
\,\,\,$b(\bar\psi \psi)^4$ \,\,\,  to the above  
Lagrangian which in this model become pure three-
nucleon-and four-nucleon-forces respectively. A modification of the
Walecka model, including higher orders of many-body forces,
as suggested by Zimanyi and Moskowski [3,4]  also 
becomes equivalent to ZRM for infinite nuclear matter in MFA if
one replaces \,$G_{s}^{2}$ in Eq. (3) by  
$G_{s}^{2}\,/\,(\,1+\, (\bar\psi \psi)/M\,)$. 

To illustrate still further the risks of associating chiral symmetry
invariance to the Walecka model we show below that within 
MFA one can not
even distinguish if this symmetry is realized in a non-linear [1] or
linear form. We start with the well-known 
Nambu-Jona-Lasinio(NJL) model [5], which has a linear realization 
of chiral symmetry. We include a vector interaction 
in a current-current form and the Lagrangian remains invariant under
a linear chiral transformation of the baryon field $\psi$
\\
$$ {\cal L}_{LC} = \bar \psi i \gamma_{\mu} \partial^{\mu } \psi 
 + \,\frac{1}{2}\,G_{s}^{2}\,\,(\,\, (\bar\psi \psi)^2 \,\, 
 + (\bar \psi i \gamma_{5}\vec{\tau} \psi)^2\,\,)
 \,\,\,-\frac{1}{2}\,G_{v}^2  
(\bar \psi \gamma_{\mu }\psi)^2\,\,
.\,\,\,,\eqno(4)$$
\\ 
For static, infinite nuclear matter, the three-vector momentum
and spin dependent interaction average to zero due to rotational 
symmetry and the Lagrangian reduces to  
\\
$$ {\cal L}_{LC} = \bar \psi i \gamma_{\mu} \partial^{\mu } \psi 
\,\,\,  +\,\frac{1}{2}\,G_{s}^{2} (\bar\psi \psi)^2 - \,
\frac{1}{2}\,G_{v}^2  
(\bar \psi \gamma_{\mu }\psi)^2,\,\,\,,\eqno(5)$$
\\
which is identical to that of Eq. (3) 
, except for the baryonic mass term.
In the MFA, we linearize the interaction in Eq. (3) by closing
the Fermi loop. It means replacing $\,\,(\bar \psi\,\Gamma_{\alpha}\,\psi)^2$
by 2$\,\bar \psi\,\Gamma_{\alpha}\,\psi
\langle {\bar\psi}\,\Gamma_{\alpha}{\psi}\rangle$. In 
nuclear matter we have only $\Gamma_{\alpha}$\,=\,
1 and $\gamma_o$. Here $\langle {\bar\psi}\,\Gamma_{\alpha}{\psi}\rangle$
is the vacuum (ground state) expectation value of the operators. For 
a Lorentz-invariant and parity-conserving vacuum, the only non-vanishing
term is $\langle {\bar\psi}{\psi}\rangle$.

So, in the same spirit of the dynamical quark mass generation mechanism
of the NJL model, we can associate the nucleon rest mass $\,M \,$ to 
\\
$$ M\,\,=\,-G_{s}^2 \langle {\bar\psi}{\psi}\rangle_{vac}=
\,\, \frac{G_{s}^{2}M \gamma}{(2\pi)^{3}}\int_{0}^
{\Lambda}{ \frac{d^{3}k}{E(k)}}\,\eqno(6)$$  
\\
therefore obtaining a mean field Lagrangian identical to the ZRM
model in MFA 
\\
$$ {\cal L} = \bar \psi i \gamma_{\mu} \partial^{\mu } \psi -
(M\,-\,G_{s}^{2}\langle {\bar\psi}{\psi}\rangle)
\bar \psi  \psi
   - \,G_{v}^2  
(\bar \psi \gamma_{o}\psi)\langle {\bar\psi}\gamma_{o}{\psi}\rangle
,\,\,\,,\eqno(7)$$
\\
where now $\langle {\bar\psi}\,\Gamma_{\alpha}{\psi}\rangle$
means the expectation value of these operators in the nuclear matter
ground state (all the nucleon states filled up to Fermi momentum) and
$\gamma$ stands for the degeneracy factor equal to four.
By looking at the problem in this perspective we have
introduced a constraint, as in the standard 
NJL model. Thus, in order to have a
nontrivial nucleon mass solution for the Eq. (6), 
the coupling constant $G_{s}^2$
must be greater than a critical value $G_{crit}\,=\,
(4\pi^2)/(\gamma\,\Lambda^2)$. Here $\Lambda$ is the cut-off which
fixes the bare nucleon mass M given by the gap equation
\\
$$ 1\,\,-\,\,C_{s}^{2}
\,\,\frac{\gamma}{2\pi^{2}}\int_{0}^{\Lambda/M}\frac{ x^2dx}
{\sqrt(1+x^2)}\,\,\,\,=0\,\eqno(8)$$  
\\
where we have identified $G_{s}^2$ to the Walecka coupling
constant $C_{s}^{2}/M^2$ and $\,x\,=\,k/M$\,
is a dimensionless variable. By fixing\,\,
M=938.27 MeV\,\, and the value of $C_{s}^2$ which gives the
correct nuclear matter saturation properties [6,7]
we have obtained $\Lambda\,= \,328.5$MeV. 
Then,
Eq.(7) will give also the same equation of state of the Walecka
model, but now the chiral symmetry is 'realized' in a linear 
way. We are aware that our drawback to this chiral approach
is the existence of a zero-frequency mode, the Nambu-Goldstone
boson, which in this case is a pseudoscalar-isovector 
nucleon-antinucleon mode that we can not identify to the pion. 
 
In summary, all the arguments we have presented 
stress the difficulties of extracting , from a nuclear matter
description at the MFA level, a 
justification of Walecka model coming from chiral effective Lagrangians. 
To be more specific, if one adds to the Weinberg
Lagrangian [8] the four-nucleon term, at the MFA level one arrives 
at a equivalent Walecka model as presented in ref. [1]. 
However, if we add any other term involving field derivatives
which explicitly breaks the chiral symmetry 
we arrive also at an equivalent Walecka model. We have 
shown that, the same procedure 
may be extended to a model where the chiral symmetry can be realized 
linearly. To conclude, we believe that what is in fact behind all
of this is the fact that the MFA of Walecka model for infinite
nuclear matter gives simply the same results of ZRM.
We claim that is more appropriate and consistent to think 
just about a distorted picture furnished by MFA than to claim any
chiral realization of hadronic phenomenological model. In this 
approximation, the pion which realizes the non-linear chiral symmetry
can not couple to the nucleons. So, the only way to conclude something
about chiral effective Lagrangians describing bulk nuclear 
matter properties is to go beyond this approximation, where
the identification with the Walecka model description will be
lost. 
\newpage
\paragraph{References}
\begin{description}
\item[1 -] Graciela Gelmini,Bruce Ritzi, Phys.Lett.{\bf{B357}},431(1995)
\item[2 -] J.D.Walecka, Ann.Phys.{\bf{83}},491(1974); B.D. Serot and J.D.
Walecka, Adv. in Nucl. Phys. vol.16, (Plenum, N.Y. 1986)
\item[3 -] J.Zimanyi and S.A.Moszkowski, Phys.Rev.{\bf{C42}},1416 (1990)
\item[4 -] A.Delfino,C.T.Coelho and M.Malheiro, Phys.Rev.{\bf{C51}},2188(1995)
\item[5 -] Y.Nambu and G. Jona-Lasinio, Phys.Rev.{\bf{122}}(1961)245;\,{\bf{124}}
(1961)246.
\item[6 -] R.J.Furnsthal and B.D.Serot,Phys.Rev.{\bf{C41}},(1990)1416
\item[7 -]  A.Delfino, C.T.Coelho, and M.Malheiro,
Phys.Lett.{\bf{B345}}361(1995)
\item[8 -]  S. Weinberg, Phys.Rev.Lett.{\bf{18}},(1967)188;Phys.Rev.{\bf{166}}
(1968)1568;Physica {\bf{A96}}(1979)372
\end{description} 
\end{document}